\documentstyle[11pt,epsfig,amssymb]{article}
\title{\bf  Quark-hadron phase transition in massive gravity}
\author{K. Atazadeh\thanks{email: atazadeh@azaruniv.ac.ir}
\\{\small Department of Physics, Azarbaijan Shahid Madani University, Tabriz 53741-161, Iran}}
\begin{document}
\maketitle
\begin{abstract}
We study the quark–-hadron phase transition  in the framework of massive gravity. We show that the modification of the FRW cosmological equations leads to the quark–-hadron phase transition  in the early  massive Universe. Using numerical analysis, we consider that a phase transition based on the chiral symmetry breaking after the electroweak transition, occurred at approximately $10 \mu$ seconds after the Big Bang to convert a plasma of free quarks and gluons into hadrons.
 \end{abstract}

\section{Introduction}

In the expanding and cooling standard cosmology, the early Universe
experiences a series of symmetry-breaking phase transitions, resulted of the topological defects forming.
Study such phase transitions help us to  understand better the evolution of the
early Universe, characterized by the existence of a quark-gluon plasma phase transition.
In this way, we concentrate on possible scenarios which allow such a
phase transition occurrence. Here, we follow \cite{harko} which have considered the
quark-gluon phase transition in a cosmologically transparent perspective. It is worth mentioning that QCD predicts
the existence of phase transition from the quark-gluon plasma phase to hadron gas one.
The phase transition in QCD which depends on
singular behavior of the partition function, can be described by a first or second order phase transition, and also it can be outlined by a crossover phase transition, that depends on the values of the quark masses. First time, the possibility of a phase transition in the gas of quark-gluon bags was studied in \cite{Gorenstein}. The results of the first, second
and higher order transitions have been  presented in literatures. Furthermore, the possibility of no phase transitions was mentioned in
\cite{Greiner}. Lattice QCD calculations carried out for two quark flavors propound that QCD makes a
smooth crossover transition at a temperature of $T_c \sim150$ MeV \cite{Aoki}. In the early Universe crossover phase transition may
be responsible for the formation of relic quark–-gluon objects which may have
survived. In this work, we study the phase transition under certain conditions that a gas of extended hadrons could
generate phase transitions of the first or second order, and also we consider a smooth crossover transition which is qualitatively same as the lattice QCD.

The color deconfined quark--gluon plasma cooling down, below the critical temperature
believed to be around $T_c\sim 150$ MeV leads to form color confined hadrons
(mainly pions and a tiny amount of neutrons and protons, since the net baryon number should be
conserved). Therefore, such a new phase does not assemble quickly. Actually, a first order
phase transition requires some supercooling to produce the energy used in composing the surface of the
bubble and the new hadron phase. In \cite{K} a first order quark-hadron phase transition in the
expanding Universe has been expressed. During the hadron bubble nucleating, hidden heat is
blurted and a spherical shock wave propagates to the surrounding supercooled quark–gluon plasma.
The plasma thus formed is reheated and reaches the critical temperature, preventing further
nucleation in a region shifted by means of one or more shock fronts. Generally, bubble growth is explained by
deflagrations where a shock front precedes the actual transition front. The ending of the nucleation
occurs when the whole Universe has reheated $T_c$. The prompt stopping of this phase transition, in
about $0.05 \mu$s, contributes the cosmic expansion completely insignificant over this period.
Finally, the transition ends when all quark-gluon plasma has been reformed into hadrons surrender of possible quark
nugget production. The quark-hadron phase transition and its
cosmological implications have been widely studied in the context of general relativistic
cosmology in \cite{quark}-\cite{ata}.

In this paper we study quark-gluon phase transition in the framework of a covariant massive gravity  model recently proposed in \cite{derham10,derham11}. In order to have a consistent theory, nonlinear terms should be tuned to remove order by order the negative energy state in the spectrum \cite{boulware72}. The model under consideration follows from a scheme originally studied in \cite{arkani03,creminelli05} and at the complete nonlinear level with an arbitrary reference metric has been not found any ghosts \cite{Hassan:2011tf,Hassan:2011ea}.
The treated theory exploits several important features. Actually, the graviton mass typically demonstrates itself on cosmological scales at late times thus providing a natural explanation of the late lime acceleration of the Universe \cite{car2012}. Additionally, the theory has exotic solutions in which the graviton mass affects the dynamics at early times. We show that in the massive gravity theory, it is possible to have cosmological quark-gluon phase transition and it exists even for spatially flat (i.e. ${\cal K}=0$) cosmological models.

\section{Massive cosmological equations}\label{massive}

We review  cosmological equations of the massive gravity introduced in \cite{derham10,derham11}. According to  the formalism used in \cite{massive}, we define a four-dimensional pseudo-Riemannian manifold $({\cal M},g)$ and the dynamics can be determined by the action as follows
\begin{equation}\label{totaction}
S=-\frac{1}{8\pi G}\int \left(\frac{1}{2} R+m^2  \mathcal{U}\right) d^{4}x \ + \ S_{m}, \nonumber
\end{equation}
where $G$, $R$ and $S_m$ are the Newton gravitational constant, the Ricci scalar and ordinary matter field action, respectively. The mass coupled potential term is defined by
\begin{eqnarray}\label{potential}
\mathcal{U}&=&\frac{1}{2} (K^2-K^\nu_\mu K^\mu_\nu)+\frac{c_3}{3!}\epsilon_{\mu\nu\rho\sigma}\epsilon^{\alpha\beta\gamma\sigma}K_\alpha^\mu K_\beta^\nu K_\gamma^\rho\nonumber\\ \nonumber
&&+\frac{c_4}{4!}\epsilon_{\mu\nu\rho\sigma}\epsilon^{\alpha\beta\gamma\delta}K_\alpha^\mu K_\beta^\nu K_\gamma^\rho K_\delta^\sigma, \nonumber
\end{eqnarray}
here $c_3$, $c_4$ are arbitrary dimensionless real constants and also $\epsilon_{\mu\nu\rho\sigma}$ is the Levi-Civita tensor density.
$$
K^\mu_\nu=\delta^\mu_\nu-\gamma^\mu_\nu,
$$
$\gamma^\mu_\nu$ being defined by the relation
$$
\gamma^\mu_\sigma\gamma^\sigma_\nu=g^{\mu\sigma} f_{\sigma \nu},
$$
with $f_{\sigma_\nu}$ that is, a symmetric tensor field. The quantity $m_g=\hbar m/c$ is called the graviton mass.

We consider the Friedmann-Lema\^{\i}tre-Robertson-Walker (FLRW) Universe with three-dimensional spatial curvature ${\cal K}=0,\pm1$, explained by the line element
\begin{eqnarray}
ds^2&=&g_{\mu\nu}dx^\mu dx^\nu\\\nonumber
&=&dt^2-a(t)^2\left[\frac{dr^2}{1-{\cal K}r^2} +r^2(d\theta^2+\sin(\theta)^2d\phi^2)\right].
\end{eqnarray}
The first Friedmann equation for generic values of the dimensionless constants $c_3$ and $c_4$ and imposing the Bianchi identities, reads
\begin{eqnarray}\label{mgfriedmann}
3\frac{\dot{a}^2+{\cal K}{a^{2}}}{a^2}&=&m^2 \left(4c_3+c_4-6+3C\frac{3-3c_3-c_4}{a}\right. \nonumber\\
&&\left.+3C^2\frac{c_4+2c_3-1}{a^2}-C^3\frac{c_3+c_4}{a^3}\right) + 8\pi G \rho~,
\end{eqnarray}
where $C$ is an integration constant. The conservation equations for the matter component is
\begin{equation}\label{fluid}
\dot{\rho}+3H(\rho+p)=0~,
\end{equation}
with $H=\dot{a}/a$. Also we define a constant equation of state parameter $w=p/\rho$.\\
Furthermore, in the consequent analysis the parameter space is reduced to the subset $c_3=-c_4$ \cite{koyama11}, that is the simplest choice that presents a successful Vainshtein effect in the weak field limit.

We can rewrite equation (\ref{mgfriedmann}) as follows
\begin{eqnarray}\label{friedmann}
H^{2}=\frac{\kappa}{3}\rho - \frac{\mathcal{K}}{a^2}+ \frac{m^2}{3} \left( A_1+\frac{A_2}{a}+\frac{A_3}{a^2 }\right),
\end{eqnarray}
where $\kappa=8 \pi G$,  $a/C \rightarrow a$ and
\begin{eqnarray}
A_1=-3c_4-6,~~~~~~~~~~
A_2=3\left(3 +2c_4 \right),~~~~~~~~~
A_3&=&-3\left(1 +c_4 \right).
\end{eqnarray}

\section{Quark--hadron phase transition}
In this section, we interpret the appropriate physical quantities of the quark--hadron phase transition,
which will be used in the following sections in the framework of the massive gravity.
We know that at the phase transition the scale of the cosmological QCD transition is given by the Hubble radius $H^{-1}$, which is $H^{-1} \sim M_4/T ^{2}_{c}\sim 10$ km, here $T_c$ is the critical temperature. Inside of
the Hubble volume has the mass about $\sim 1M_{\odot}$. The timescale of QCD is $1$ fm/c $\approx 10^{-23}$ s which should be compared
with the expansion time scale, $10^{-5}$ s. Even the rate of the weak interactions passes the
Hubble rate by a factor of $10^{7}$. As a result, in this phase photons, leptons, quarks and gluons (or
pions) are lightly coupled and may be characterized as a single, adiabatically expanding fluid [12].
In consideration of the quark--hadron phase transition it is essential to determine the equation of
state of the matter, in both hadron and quark state. Specifying an equation of state is equivalent to
give the chemical potential $\mu$ and the pressure as a function of the temperature $T$. The quark chemical potentials
at high temperatures are equal, because weak interactions keep them in chemical
equilibrium, and the chemical potentials for leptons are zero. So the chemical
potential for a baryon is characterized by $\mu_B = 3\mu_q$ . The baryon number density of an ideal Fermi gas
of three quark flavors is defined by $n_B \approx T ^{2}\mu_B/3$ which leads to $\mu_B/T \sim10^{-9}$ at $T >T_c$. At low
temperatures $\mu_B/T \sim 10^{-2}$. Thus, an excellent approximation
for the study of the equation of state of the cosmological matter in the early Universe, is assumption of a vanishing chemical potential at the
phase transition temperature in both quark and hadron phase. In
addition to the strongly interacting matter, we suppose that in each phase there are present leptons
and relativistic photons, obeying equations of state similar to that of hadronic matter [6].
The equation of state of the matter in the quark phase can be have the following form
\begin{equation}\label{eq3.1}
\rho_q=3a_qT^{4}+V(T),~~~~~~~~~~~~p_q=a_qT^{4}-V(T),
\end{equation}
where $a_q = (\pi^{2}/90)g_q $, with $g_q = 16 + (21/2)N_F + 14.25 = 51.25$ and $N_F = 2$. $ V (T )$ is the
self-interaction potential. For $V(T)$ we use the expression in [19]
\begin{equation}\label{eq3.2}
V(T)=B+\gamma_T T^{2}-\alpha_{T}T^{4},
\end{equation}
where $B$ is the bag pressure constant,  $\alpha_T = 7\pi ^{2}/20$ and $\gamma_T = m^{2}_s/4$
with $m_s$ the mass of the strange quark in the range $m_s \in(60–-200)$ MeV. The potential form $V$ corresponds to a physical
model in which the quark fields are interacting with a chiral field constructed with a scalar field and the $\pi$ meson
field. By ignoring the temperature effects, the equation of state in the
quark phase is in the form of $p_q = (\rho_q- 4B)/3$, in the MIT bag model. Results
gained in low energy hadron spectroscopy, heavy ion collisions and phenomenological fits of
light hadron properties give $B^{1/4}$ between $100$ and $200$ MeV [25].
In the hadron phase, we take the cosmological fluid which consists of an ideal gas of massless
pions and of nucleons explained by the Maxwell–-Boltzmann statistics, with energy density $\rho_h$
and pressure $p_h$, respectively. The equation of state can be written approximately as
\begin{equation}\label{eq3.3}
p_h(T)=\frac{1}{3}\rho_h(T)=a_\pi T^4,
\end{equation}
where $a_\pi =(\pi^2/90)g_h$ and $g_h=17.25$.

The critical temperature $T_c$ is defined by the condition $p_q(T_c) = p_h(T_c)$ [6], and in
the present model given by
\begin{equation}\label{eq3.4}
T_c=\left(\frac{\gamma_T+\sqrt{\gamma_T^{2}+4B(a_q+\alpha_T-a_\pi)}}{2(a_q+\alpha_T-a_\pi)}\right)^{1/2}.
\end{equation}
For $m_s = 200$ MeV and $B^{1/4} = 200$ MeV the transition temperature is of the order $T_c \approx 125$ MeV. When the phase transition is of first order, all the physical quantities, like the energy
density, pressure and entropy have discontinuities behavior across the critical curve. The ratio of the
quark and hadron energy densities at the critical temperature, $\rho_q (T_c)/\rho_h(T_c)$, is of the order of
$3.62$ for $m_s = 200$ MeV and $B^{1/4} = 200$ MeV. If the temperature effects in the self-interaction
potential $V$ are neglected, $\alpha_T = \gamma_T \approx 0$, thus we get the well-known relation between the
 bag constant and critical temperature, $B = (g_q -g_h)\pi^2T^4_
c /90$ [6].

\begin{figure}
\begin{center}
\epsfig{figure=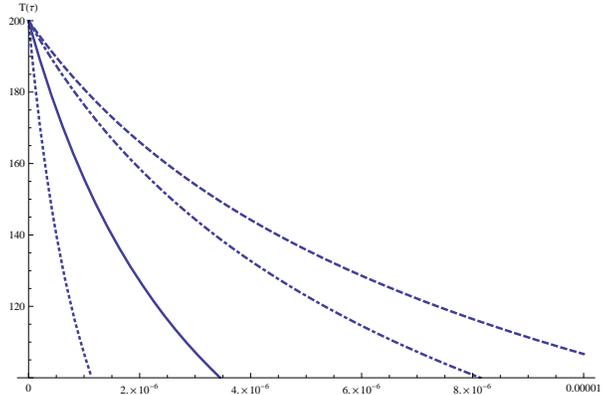,width=8cm}
\end{center}
\caption{\footnotesize  Variation of $T(\tau)$ as a function of
time ($\tau$) for different values of $c_4$: $c_4= 8\times10^{8}$
 (solid curve), $c_4=8\times 10^{9}$
 (dashed curve), $c_4=2\times 10^{10}$
(dotted curve) and $c_4=1\times 10^{11}$
 (dotted-dashed curve). We have taken $B^{1/4}=200$ MeV, $\kappa=1$ and ${\cal K}=-1$.}
\end{figure}

\section{Dynamical interpretation of the massive Universe during the quark--hadron phase transition}
The quantities to be followed through the quark-hadron phase
transition in the massive gravity are the temperature $T$, the energy
density $\rho $ and the scale factor $a$.
Note that these quantities are determined by the gravitational field
equations (\ref{friedmann}) and (\ref{fluid}) and by the equations of
state (\ref{eq3.1}), (\ref{eq3.2}) and (\ref{eq3.3}). We must study
now the evolution of the massive Universe before, during and after the
phase transition.

Before the phase transition, $T>T_{c}$ the massive Universe is in the
quark phase. With the use of the equations of state of the quark
matter, and the conservation equations for the matter component, equation (\ref{fluid})
can be written in the form
\begin{equation}\label{eq4.1}
H=\frac{\dot{a}}{a}=-\frac{3a_{q}-\alpha
_{_{T}}}{3a_{q}}\frac{\dot{T}}{T}-\frac{1}{6}\frac{\gamma
_{_{T}}}{a_{q}}\frac{\dot{T}}{T^{3}},
\end{equation}
by integrating we have scale
factor-temperature relation as
\begin{equation}\label{eq4.2}
 a(T)=a_{0}~T^{\frac{\alpha _{_{T}}-3a_{q}}{3a_{q}}}\exp \left(
\frac{1}{12}\frac{\gamma _{_{T}}}{a_{q}}\frac{1}{T^{2}}\right) ,
\end{equation}
where $a_{0}$ is a constant of integration.

By using  equation (\ref{eq4.1}) we have an expression, describing the evolution of the
temperature of the massive Universe in the quark phase, given by
\begin{eqnarray}\label{eq4.3}
\frac{dT}{d\tau}&=&\frac{-T^3}{U_0 T^2+U_1}\times
 \\\nonumber&&\sqrt{\frac{1}{3} m^2 \left(\frac{A_3 T^{-2\xi} e^{-\frac{\eta}{T^2}}}{a_0^2}+\frac{A_2
   T^{-\xi} e^{-\frac{\eta}{2 T^2}}}{a_0}+A_1\right)+\frac{1}{3} \kappa  \left(3 a_q T^4+B- \alpha _{T}T^4+\gamma _{T}T^2
   \right)-\frac{{\cal K} T^{-\xi} e^{-\frac{\eta}{T^2}}}{a_0^2}},
\end{eqnarray}
where we have denoted
\begin{eqnarray}\label{eq4.4}
U_{0}=1-\frac{\alpha_{_{T}}}{3a_{_{T}}},
~~~~~~~
\eta=\frac{\gamma _T}{6a_q},
~~~~~~~~
\xi=\frac{\alpha _T-3 a_q}{3 a_q},
~~~~~
U_{1}=\frac{\gamma_{_{T}}}{6a_{q}}.
\end{eqnarray}
We solve equation (\ref{eq4.3}) numerically and the
result is presented in Figure 1 which shows the behavior of
temperature as a function of cosmic time $\tau$ in a massive Universe filled with quark matter for different values of
$c_4$. It can be seen that the temperature drops faster for higher values of $c_4$.

\subsection{Self-interaction potential $(V(T)=B)$}
Let us take attention on the phase transition era, namely the simple case where
temperature corrections can be neglected in the self-interaction
potential $V$. Then $V=B={\rm cons.}$ and equation of state of the
quark matter is given by that of the bag model, $p_{q}=\left( \rho
_{q}-4B\right) /3$. Equation (\ref{fluid}) may then be integrated
to give the scale factor massive Universe as a function of temperature
\begin{equation}\label{eq4.5}
a(T)=\frac{c}{T},
\end{equation}
where $c$ is a constant of integration.

\begin{figure}
\begin{center}
\epsfig{figure=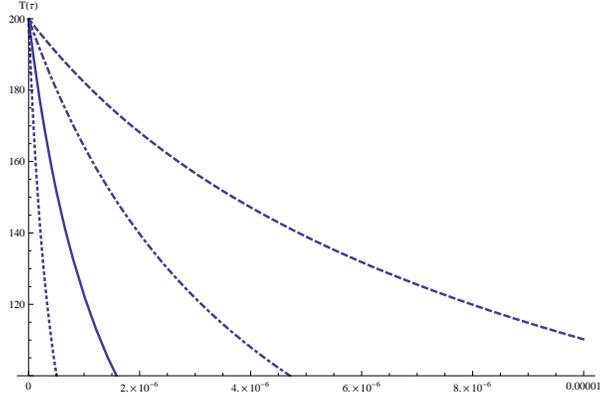,width=8cm}
\end{center}
\caption{\footnotesize  Variation of $T(\tau)$ as a function of
time ($\tau$) for different values of $c_4$: $c_4= 8\times10^{8}$
 (solid curve), $c_4=8\times 10^{9}$
 (dashed curve), $c_4=2\times 10^{10}$
(dotted curve) and $c_4=1\times 10^{11}$
 (dotted-dashed curve). We have taken $B^{1/4}=200$ MeV, $\kappa=1$ and ${\cal K}=-1$.}
\end{figure}
Using equations (\ref{fluid}) and (\ref{friedmann}), the time
dependence of temperature can be written as follows
\begin{equation}\label{eq4.7}
\frac{1}{T}\frac{dT}{d\tau}=-\sqrt{
\frac{\kappa}{3}\left(3a_qT^{4}
+B\right)-{\cal K}T^{^2}+\frac{m^{2}}{3}\left(A_{1}+A_{2}T+
A_{3}T^{^2}\right)}.
\end{equation}
The equation (\ref{eq4.7}) can be solved numerically and plotted in figure 2 which shows the
behavior of temperature as a function of cosmic time $\tau$ in a massive Universe filled with quark matter
for different values of $c_4$.
\subsection{Formation of hadrons}
In the phase transition regime, pressure and temperature are
constant and quantities like the  entropy $S=sa^{3}$ and enthalpy
$W=\left( \rho +p\right) a^{3}$ are conserved. Following
\cite{harko,quark}, we substitute  $\rho \left( \tau\right) $ by
$h(\tau)$, the volume fraction of matter in the hadron phase, by
defining
\begin{equation}\label{eq4.11}
\rho \left( \tau\right) =\rho _{_{H}}h(\tau)+\rho _{_{Q}}\left[ 1-h(\tau)\right]
=\rho _{_{Q}} \left[ 1+nh(\tau)\right] ,
\end{equation}
where $n=\left( \rho _{_{H}}-\rho _{_{Q}}\right) /\rho _{_{Q}}$.
Start of the phase transition is described by
$h(\tau_{c})=0$ where $\tau_{c}$ is the time representing the
beginning of the phase transition and $\rho \left( \tau_{c}\right)
\equiv \rho _{_{Q}}$, during the end of the transition is
characterized by $h\left( \tau_{h}\right) =1$ with $\tau_{h}$
being the time signalling the end and corresponding to $\rho \left(
\tau_{h}\right) \equiv \rho _{_{H}}$. For $\tau>\tau_{h}$ the
Universe goes to the hadronic phase.

Now, equation (\ref{fluid}) gives
\begin{equation}\label{eq4.12}
\frac{\dot{a}}{a}=-\frac{1}{3}\frac{\left( \rho _{_{H}}-\rho
_{_{Q}}\right) \dot{h} }{\rho _{_{Q}}+p_{c}+\left( \rho _{_{H}}-\rho
_{_{Q}}\right) h}=-\frac{1}{3}\frac{r \dot{h}}{1+rh},
\end{equation}
where we have implied $r=\left( \rho _{_{H}}-\rho _{_{Q}}\right)
/\left( \rho _{_{Q}}+p_{c}\right)$. From the above equation we can
obtain the relation between the scale factor and the
hadron fraction $h(\tau)$
\begin{equation}\label{eq4.13}
a(\tau)=a\left( \tau_{c}\right) \left[ 1+rh(\tau)\right] ^{-1/3},
\end{equation}
where the initial condition $h\left(\tau_{c}\right)=0$ has been
used. Now, using equations (\ref{friedmann}) and (\ref{eq4.13}) we
get the time evolution of the matter fraction in the hadronic
phase
\begin{eqnarray}\label{eq4.14}
\frac{dh}{d\tau}&=&-3\left( h+\frac{1}{r}\right)\times\\\nonumber&&
\sqrt{
\frac{\kappa\rho _{_{Q}}\left[ 1+nh(\tau)\right]}{3} -\frac{{\cal K}\left[ 1+rh(\tau)\right]}{a^{2}\left( \tau_{c}\right)}  ^{2/3}+\frac{m^{2}}{3}\left(A_{1}+\frac{A_{2}\left[ 1+rh(\tau)\right] ^{1/3}}{a\left( \tau_{c}\right)} +
\frac{A_{3}\left[ 1+rh(\tau)\right] ^{2/3}}{a^{2}\left( \tau_{c}\right)} \right)}.
\end{eqnarray}
Figure 2 shows variation of the hadron fraction $h(\tau)$ as a
function of $\tau$  for different values of $c_4$. The higher values of $c_4$ end the phase transition in shorter period of time.
\begin{figure}
\begin{center}
\epsfig{figure=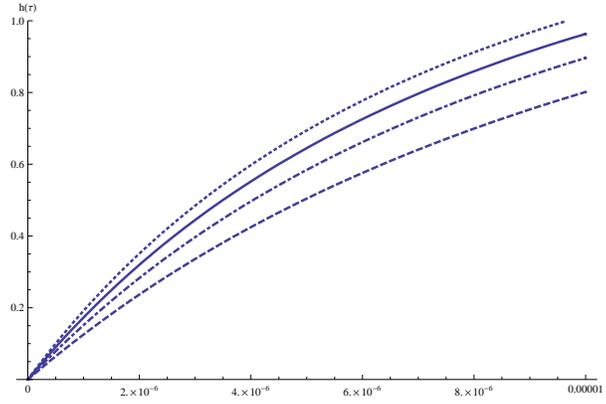,width=8cm}
\end{center}
\caption{\footnotesize  Variation of $h(\tau)$ as a function of
time ($\tau$) for different values of $c_4$: $c_4= 8\times10^{8}$
 (solid curve), $c_4=8\times 10^{9}$
 (dashed curve), $c_4=2\times 10^{10}$
(dotted curve) and $c_4=1\times 10^{11}$
 (dotted-dashed curve). We have taken $B^{1/4}=200$ MeV, $\kappa=1$ and ${\cal K}=-1$.}
\end{figure}

\subsection{Pure hadronic era}
The scale factor of the Universe at the end of the phase transition has the value
\begin{equation}
a\left( \tau_{h}\right)
=a\left( \tau_{c}\right) \left( r+1\right) ^{\frac{-1}{3}}.
\end{equation}
Also, the energy density of the pure hadronic matter after the phase
transition is $\rho _{h}=3p_{h}=3a_{\pi }T^{4}$. The conservation
equation (\ref {fluid}) gives
$a(T)=a\left( \tau_{h}\right) T_{c}/T$. The temperature
dependence of the massive Universe in the hadronic phase is
governed by the equation
\begin{equation}\label{eq4.19}
\frac{dT}{d\tau}=-T\sqrt{
\kappa a_\pi T^{4}-\frac{{\cal K}}{a^{2}\left( \tau_{h}\right) T_{c}^{2}}T^{2}+\frac{m^{2}}{3}\left(A_{1}+\frac{A_{2}}{a\left( \tau_{h}\right) T_{c}}T+\frac{A_{3}}{a^{2}\left( \tau_{h}\right) T_{c}^{2}}T^{2}\right)}.
\end{equation}
Variation of temperature of the hadronic fluid filled massive
Universe as a function of $\tau$ for different values of
$c_4$ is depicted in figure 3. The hadronic fluid cools down faster for higher values of $c_4$.
\begin{figure}
\begin{center}
\epsfig{figure=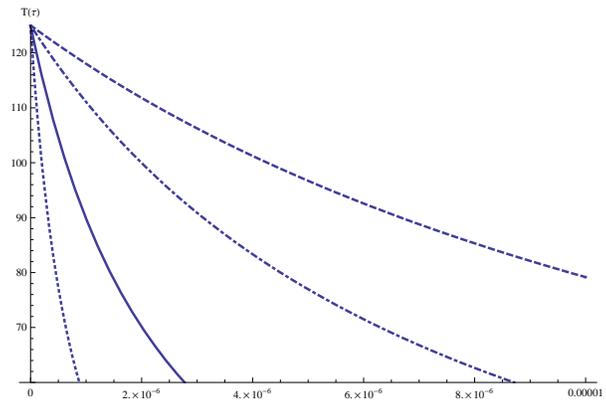,width=8cm}
\end{center}
\caption{\footnotesize  Variation of $T(\tau)$ as a function of
time ($\tau$) for different values of $c_4$: $c_4= 8\times10^{8}$
 (solid curve), $c_4=8\times 10^{9}$
 (dashed curve), $c_4=2\times 10^{10}$
(dotted curve) and $c_4=1\times 10^{11}$
 (dotted-dashed curve). We have taken $B^{1/4}=200$ MeV, $\kappa=1$ and ${\cal K}=-1$.}
\end{figure}
\section{Hight and low temperature equation of state in lattice QCD }

QCD phase transition is an important character of the particle
physics in which it is more relevant to any studied mechanisms responsible
for the evolution of the early Universe. In such a
scenario, under a crossover transition a soup of quarks and gluons interact and undergo
to form hadrons. To study the Universe at early times within
the context of massive gravity in the lattice QCD, we must
have a short review of the basic conceptions before using the results achieved in such a phase transition.
Lattice QCD is a new access that allows one to systematically
consider the non-perturbative regime of the QCD
equation of state. The QCD equation
of state was calculated by supercomputers on the lattice in \cite{66} with a heavier strange quark and two light quarks on a $(N_t = 6)×32^{3}$
size lattice. See \cite{66,69--71} for recent results in lattice QCD at high temperature.
We see that during the high temperature regime, radiation like behavior in the region at and below the critical
temperature $T_c (\approx200$ MeV) of the de-confinement transition changes remarkably.
We will see in the next section this change in behavior will also be relevant for cosmological observables.
For high temperatures interval $2.82 (100$ MeV) and $7.19 (100$ MeV) data fitting give a simple form of equation of state as
\begin{eqnarray}\label{eqq1}
\rho_{_{High}}(T) \approx \alpha T^{ 4},\\\nonumber
p_{_{High}}(T ) \approx \sigma T ^{4}.
\end{eqnarray}
By using a least squares fit, one can get $\alpha = 14.9702 \pm 009997$ and $\sigma = 4.99115 \pm 004474$ \cite{66}.
While for times before the phase transition the lattice data
match the radiation behavior very well, for times corresponding
to temperatures above $T_c$ the behavior of the lattice
data changes towards that of the matter dominated phase. We
notice that lattice studies on QCD show that the phase transition
is actually a crossover transition.

In addition to, there is an other approach to the
low temperature equation of state in the lattice QCD named Hadronic Resonance Gas model (HRG)in which QCD in the confinement
phase is considered as an non-interacting gas of fermions
and bosons \cite{72}. The conception of the HRG model is to basically
account for the strong interaction in the confinement phase
by looking at the hadronic resonances only, since these are
implicity the relevant degrees of freedom in that phase. The
HRG model predicts a good description of thermodynamic
quantities in the transition region from high
to low temperatures \cite{76}. The trace
anomaly result in HRG can also be parameterized as \cite{78}
\begin{eqnarray}\label{eqq2}
\frac{\Theta(T)}{T^{4}}
\equiv
\frac{\rho - 3p}{T^4}= a_1T +a_2T^{ 3} +a_3T^{ 4} + a_4T ^{10},
\end{eqnarray}
with $a_1 = 4.654 GeV^{-1}$, $a_2 = -879 GeV^{-3},$ $a_3 = 8081 GeV^{-4} $ and $a_4 =-7039000 GeV^{-10}$.

The calculation of the pressure, entropy
density and energy density in lattice QCD, usually arises from the
calculation of the trace anomaly $\Theta(T ) = \rho(T ) -3p(T )$.
Using the conventional thermodynamic identity, the pressure difference at
temperatures $T$ and $T_{low}$ can be written as the integral of
the trace anomaly
\begin{eqnarray}\label{eqq3}
\frac{p(T )}{T_4}-\frac{p(T_{low})}{T^{4}_{low}}=\int^{T}_{T_{low}}\frac{dT'}{T^{5}\Theta(T')}.
\end{eqnarray}
By taking the lower integration limit adequately small,
$p(T_{_{low}})$ can be neglected due to the exponential suppression and the energy density $\rho(T ) = \Theta(T ) + 3p(T )$ and
entropy density $s(T ) = (\rho+ p)/T$ can be calculated.
This technique is known as the integral method \cite{79}. Using equations
(\ref{eqq2}) and (\ref{eqq3}), we get
\begin{eqnarray}\label{eqq4}
\rho_{_{Low}}(T ) = \eta T^4 +4a_1T^{5} + 2a_2T^7 +\frac{7a_3}{4}T^8 +\frac{13a_4}{10}T^{14},
\end{eqnarray}
where $\eta =-0.112$. The trace anomaly plays an essential role
in lattice determination of the equation of state. The equation
of state is obtained by integrating the parameterizations
given in (\ref{eqq2}) over temperature as shown in (\ref{eqq3}).

\section{ Massive Universe and QCD phase transition}
\subsection{High temperature regime}
To continue, let us consider the era before the phase transition at
high temperature where the Universe is in the quark phase.
Using the conservation equation of matter and
equation of state of quark matter (\ref{eqq1}), one can write
 the Hubble parameter relation as
\begin{equation}\label{eqq5}
H = \frac{\dot{a}}{a} =-\frac{4\alpha}{3(\alpha+\sigma)}\frac{\dot{T}}{T},
\end{equation}
thus, one can solve for the scale factor
\begin{equation}\label{eqq6}
a(T ) = w_3T^{\frac{-4\alpha}{3(\alpha+\sigma)}},
\end{equation}
where $w_3$ is an integration constant.

We can write an expression to describe
the behavior of temperature of the massive Universe
as a function of time the quark phase. Using (\ref{eqq1}) and (\ref{friedmann}), we get a differential equation for the temperature as follows
\begin{equation}\label{eqq7}
\frac{dT}{dt}=-\frac{3(\alpha+\sigma)T}{4\alpha}\sqrt{\frac{\kappa}{3}\rho(T) - \frac{\mathcal{K}}{a(T)^2}+ \frac{m^2}{3} \left( A_1+\frac{A_2}{a(T)}+\frac{A_3}{a(T)^2 }\right)},
\end{equation}
here we have set $w_3=1$.
In compatible  with the HRG equation of state the lower temperature limit is $T <170 $MeV.
Equation (\ref{eqq7}) may be solved numerically and the result is
depicted in Fig. 4, which presents the behavior of  temperature
of the Universe in the quark phase as a function of cosmic
time in massive cosmology for different values of the  $c_{4}$, in the interval $282$ MeV$<T <719 $MeV in
high temperature regime. It can be seen that as the time grows
the Universe becomes cooler.
\begin{figure}
\begin{center}
\epsfig{figure=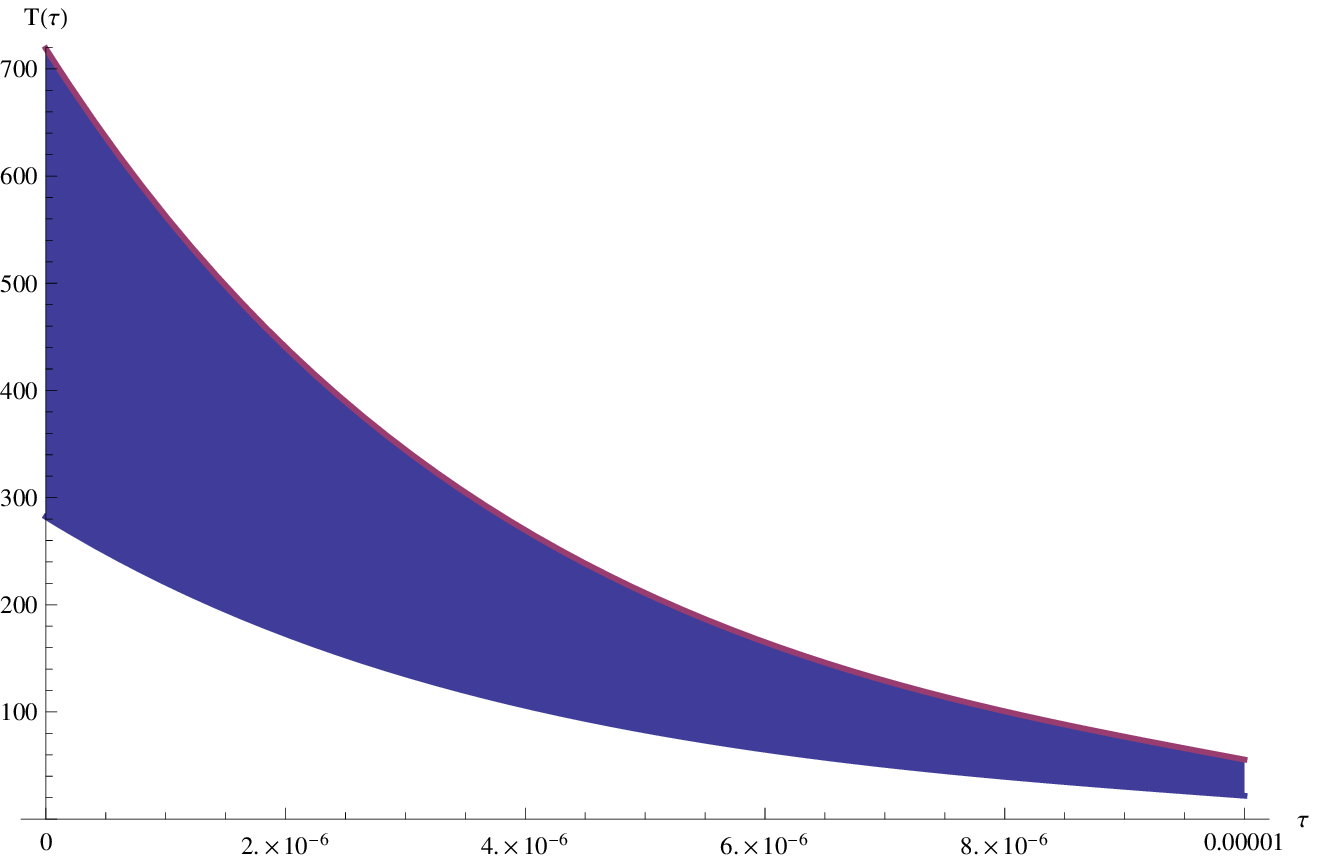,width=6cm}\hspace{5mm}
\epsfig{figure=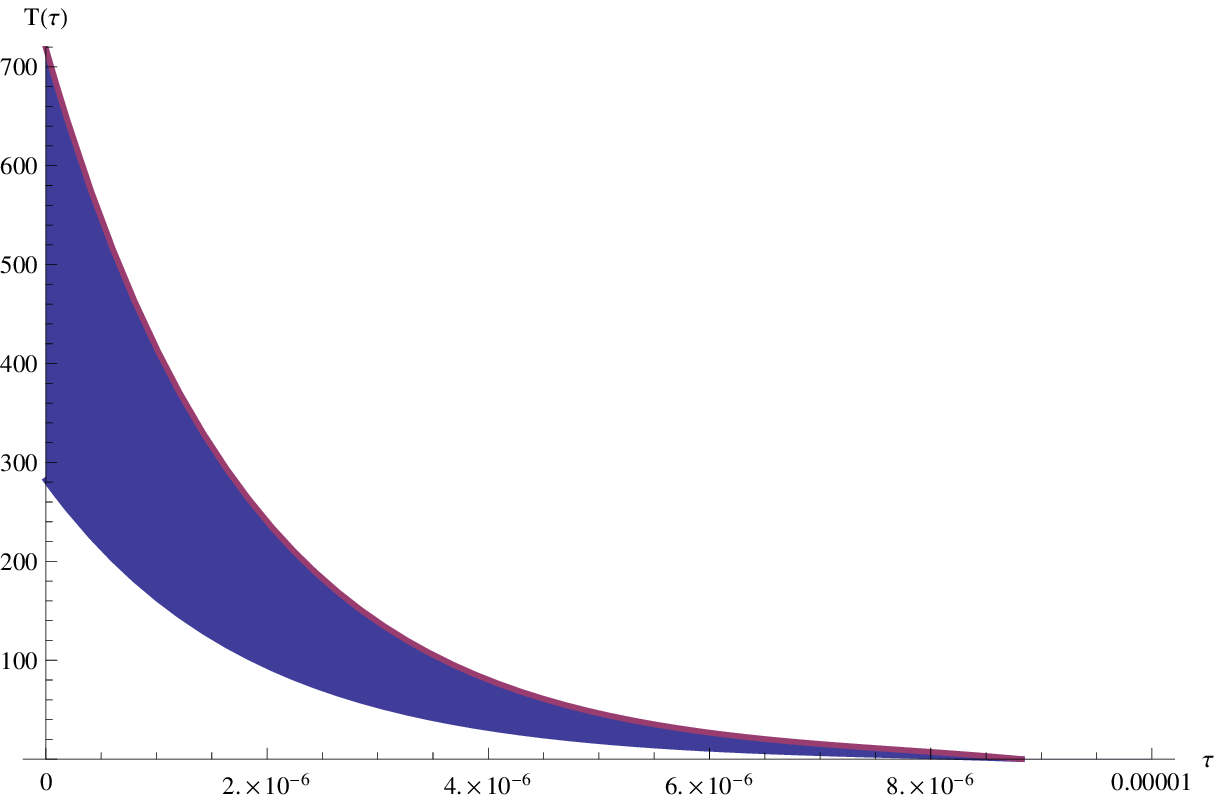,width=6cm}
\end{center}
\caption{\footnotesize The behavior of $T (\tau)$ in interval $282$ MeV $< T < 719$ MeV
as a function of $\tau$ for values of $c_4=6\times10^{10}$ (left) and $c_4=3\times10^{11}$ (right). We have taken $\kappa=1$ and ${\cal K}=-1$.}
\end{figure}

\subsection{Low temperature regime}
In this subsection we study the era before phase transition
at low temperature where the Universe is in the confinement
phase and is expressed as a non-interacting gas of fermions and
bosons \cite{72}. Using the equation of state (\ref{eqq4}) together with conservation equation of matter, we obtain the Hubble parameter as
\begin{equation}\label{eqq8}
H =\frac{\dot{a}}{a} =-\frac{12\eta T^3 +20a_1T^4 +A(T)}{3[4\eta T^4 + 5a_1T^5 +B(T)]}\dot{T},
\end{equation}
where
\begin{eqnarray}\label{eqq9}
A(T) = 14a_2T^6 +14a_3T^7 +\frac{91}{5}T^{13},\\\nonumber
B(T) =73a_2T^7 +2a_3T^8 +75a_4T^{14}.
\end{eqnarray}
We may now solve for the scale factor and obtain
\begin{equation}\label{eqq10}
a(T ) =\frac{w_4}{T (75a_1T +35a_2T^3 +30a_3T^4 +21T^{10} +60\eta)^{1/3}} ,
\end{equation}
where $w_4$ is an integration constant.
We can write an expression to explain the behavior of
temperature of the massive Universe as a function of
time in the quark phase.
By using (\ref{eqq4}) and (\ref{friedmann}), one can write a differential equation for  temperature as follows
\begin{eqnarray}\label{eqq11}
\frac{dT}{dt}=-\frac{3[4\eta T^4 + 5a_1T^5 +B(T)]T}{12\eta T^3 +20a_1T^4 +A(T)}\sqrt{\frac{\kappa}{3}\rho(T) - \frac{\mathcal{K}}{a(T)^2}+ \frac{m^2}{3} \left( A_1+\frac{A_2}{a(T)}+\frac{A_3}{a(T)^2 }\right)}
\end{eqnarray}
where we have set $w_4=1$.
We have solved equation (\ref{eqq11}) numerically, and the result is depicted
in Fig. 5, which presents the behavior of  temperature of
the Universe in the quark phase as a function of the cosmic time
for different values of $c_4$, in the interval $80$ MeV $< T < 180$ MeV at
the low temperature regime. From Fig. 5 it can be seen that in
the context of the HRG model,  $c_{4}$ enhances the cooling
process. It seems in our model the phase transition
depends on $c_4$ and  can be interpreted as
a running coupling constant for the phase transition.
\begin{figure}
\begin{center}
\epsfig{figure=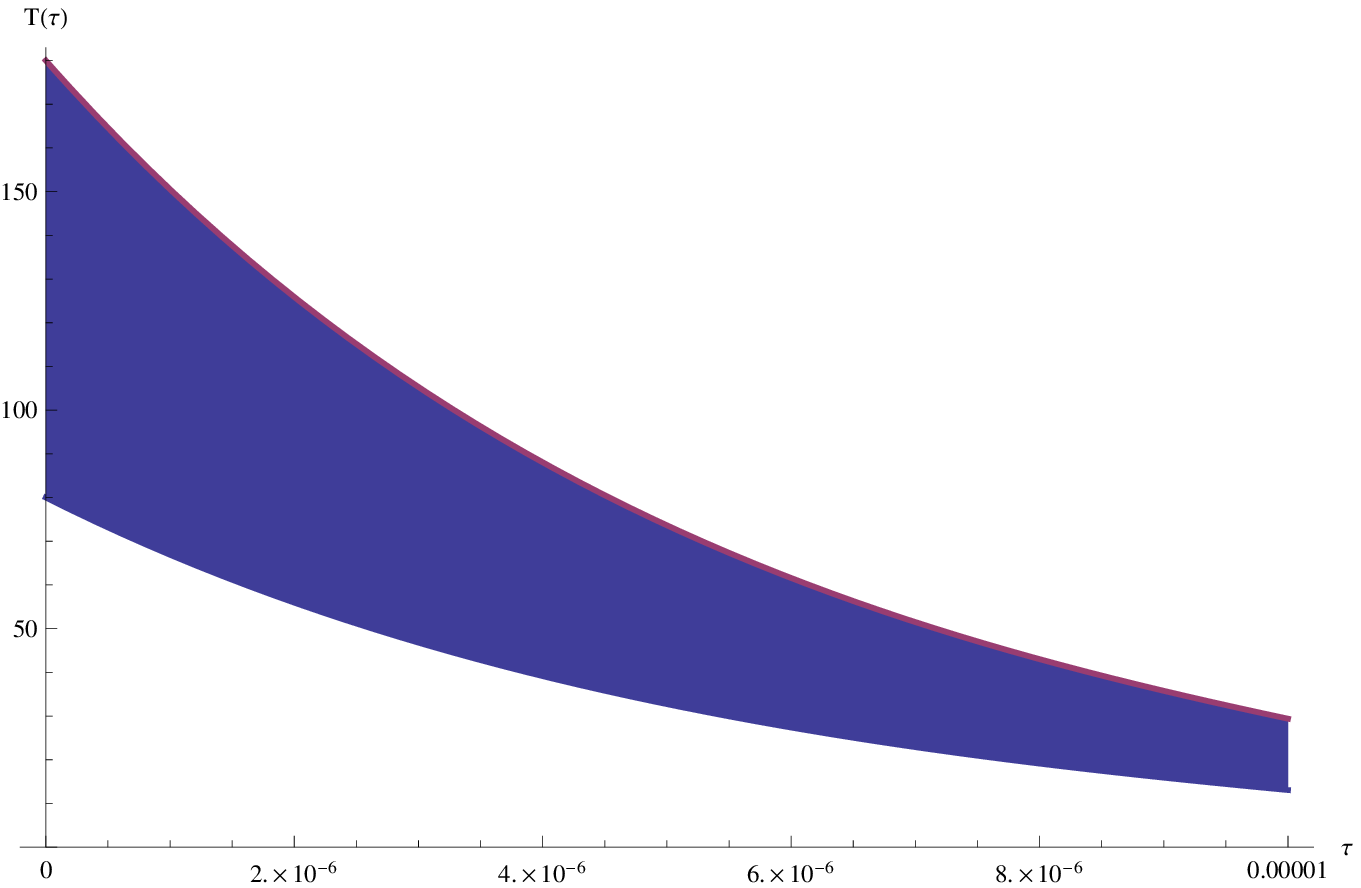,width=6cm}\hspace{5mm}
\epsfig{figure=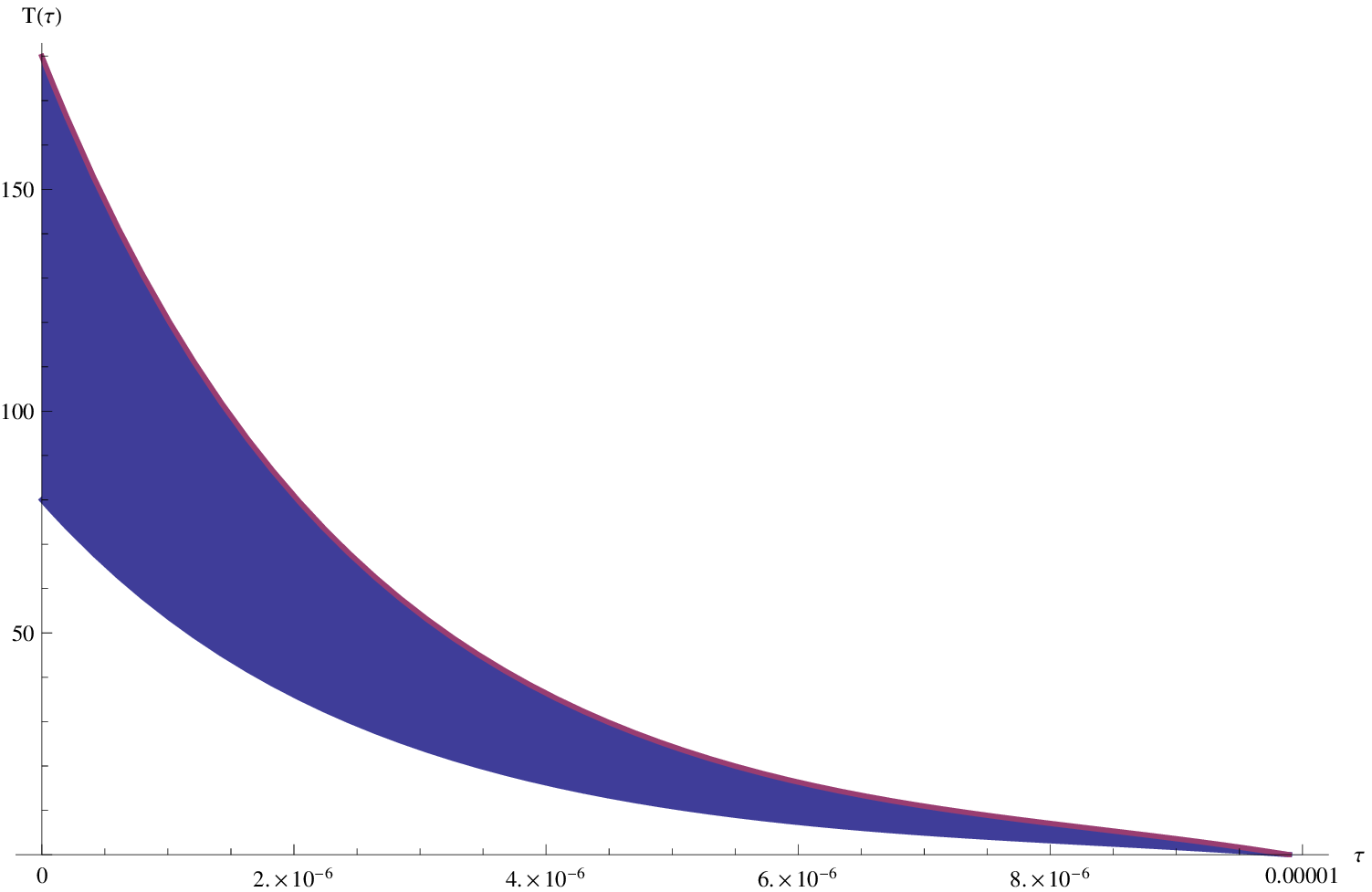,width=6cm}
\end{center}
\caption{\footnotesize The behavior of $T (\tau)$ in interval $282$ MeV $< T < 719$ MeV
as a function of $\tau$ for values of $c_4=6\times10^{10}$ (left) and $c_4=3\times10^{11}$ (right). We have taken $\kappa=1$ and ${\cal K}=-1$.}
\end{figure}

\section{Conclusion}
In this paper, we have considered the quark–-hadron phase transition
in the massive gravity, within an effective model of QCD. We have studied the time evolution of
the physical quantities in the early Universe, such as; the energy density, temperature and scale factor,
before, during, and after the phase transition.
Also, in the context of crossover transition for high and low temperatures we have discussed the time evolution of temperature and scale factor of the early Universe which is presented in section five.

We have shown that the effective temperature of the quark–gluon
plasma and of the hadronic fluid for different values of $c_4$ phase transition occurs and it decreases.
Comparing Figs. 4 and 5,
it can be seen that for the above mentioned two
models the slope of temperature, $T$ , is different during the crossover transition.
Taking into account the energy range in which the
calculations are done, one might conclude that these two approaches
to the quark–-hadron transition in the early Universe
do not predict fundamentally different ways of the evolution
of the early Universe. Note that in the massive gravity theory it is possible to have cosmological quark-gluon phase transition and it exists even for spatially flat and closed (i.e. ${\cal K}=0,+1$) cosmological models.

\section*{Acknowledgments}
 This work has been supported financially by Research Institute for Astronomy and Astrophysics of Maragha
(RIAAM) under research project NO.1/4165-58.


\end{document}